\documentclass[10pt,epsf,preprint]{aastex}

\shorttitle{The orthogonal fitting procedure for $\Sigma-D$ relations}

\shortauthors{Uro{\v s}evi{\'c} et al.}

\begin{document}

\title{The orthogonal fitting procedure for determination of the empirical $\Sigma-D$ relations for supernova remnants:
application to starburst galaxy M82}

\author{D. Uro{\v s}evi{\'c}\altaffilmark{1,2}, B. Vukoti{\'c},
\altaffilmark{3}, B. Arbutina\altaffilmark{1}, and M. Sarevska\altaffilmark{4}}

\altaffiltext{1}{Department of Astronomy, Faculty of Mathematics,
University of Belgrade, Studentski trg 16, 11000 Belgrade, Serbia;
dejanu@math.rs, arbo@math.rs}

\altaffiltext{2}{Isaac Newton Institute of Chile, Yugoslavia Branch}

\altaffiltext{3}{Astronomical Observatory, Volgina 7, 11060 Belgrade 38, Serbia;
bvukotic@aob.rs}


\altaffiltext{4}{University of Ni{\v s}, Serbia}


\keywords{galaxies: individual (M82) --- ISM:  supernova remnants --- methods: statistical --- radio continuum: ISM}

\begin{abstract}
The radio surface brightness-to-diameter ($\Sigma-D$) relation for supernova remnants (SNRs) in the starburst galaxy M82 is analyzed in a statistically more robust manner than in the previous studies that mainly discussed sample quality and related selection effects. The statistics of data fits in $\log \Sigma-\log D$ plane
are analyzed by using vertical (standard) and orthogonal regressions. As the parameter values of $D-\Sigma$ and $\Sigma-D$ fits are invariant within the estimated uncertainties for orthogonal regressions, slopes of the empirical $\Sigma-D$ relations should be determined by using the orthogonal regression fitting procedure.
Thus obtained $\Sigma-D$ relations for samples which are not under severe influence of the selection effects could be used for estimating SNR distances.
Using the orthogonal regression fitting procedure $\Sigma-D$ slope $\beta\approx3.9$ is obtained for the sample of 31 SNRs in M82.
The results of implemented Monte Carlo simulations show that the sensitivity selection effect does not significantly influence the slope of M82 relation.
This relation could be used for estimation of distances to SNRs that evolve in denser interstellar environment, with number denisty up to 1000 particles per cm$^3$.
\end{abstract}


\section{Introduction}

The relation between surface brightness $\Sigma$ and diameter $D$ for
supernova remnants (SNRs) - known as the $\Sigma-D$ relation -
is a standard way for investigating the radio brightness
evolution of these sources. In the vast majority of situations it is not feasible to observe the detailed evolution of individual SNRs over very long periods of time. However by studying the properties of samples of SNRs, which cover a range of different ages but are assumed to follow similar evolutionary paths, it is possible to analyze their statistical properties and evolution. Understanding the statistical and evolutionary properties of SNR samples and particularly using well defined samples to determine the $\Sigma-D$ relation also has an important role in providing a method of distance determination for individual SNRs. This is particularly relevant for Galactic SNRs, of which more than 200 have unknown or ill-determined distance measures (see for example, Green 2009).

In a single external galaxy all SNRs in the sample are at essentially the same distance. This makes the extragalactic samples of better quality when compared to Galactic ones, because the problems stemming from inaccurate knowledge of distances are eliminated. Additionally, Malmquist bias\footnote{The volume selection effect --  intrinsically bright objects are favored in any flux limited survey because they are sampled from a larger spatial volume.} severely acts in the Galactic samples making
them incomplete. An extragalactic sample is not {influenced by Malmquist bias. On the other hand,} the best radio instruments {at} this moment can provide detection of brighter SNRs only in the nearby galaxies. {Such a limited} survey sensitivity {results in a} selection effect {that} significantly reduces the number of detected objects within a relatively distant extragalactic system (approximately up to 15 Mpc, see Uro{\v s}evi{\' c} et al. 2005, hereafter Paper I and Chomiuk \& Wilcots 2009).

This paper presents a fitting procedure that can result, if reliable samples are used, in $\Sigma-D$ relations that are more useful in terms of
of distance estimation. The usual form of the relation is:

\begin{equation}
\Sigma=AD^{-\beta},
\end{equation}

\noindent where parameter $A$ and slope $\beta$ are obtained {by fitting of the observational data for a sample of SNRs.

The two initial empirical $\Sigma-D$ relations were derived by Poveda \& Woltjer (1968) and Milne (1970). During the 1970's and early 1980's  a number of detailed analyzes of Galactic relations were presented (eg. Clark \& Caswell 1976, Milne 1979). More critical analysis started with the work of Green (1984). A Galactic $\Sigma-D$ relation that is still quite frequently used was derived by Case \& Bhattacharya (1998). A brief review of Galactic and extragalactic relations was presented by Uro{\v s}evi{\' c} (2002). The updated Galactic $\Sigma-D$ relations were derived by Guseinov et al. (2003) and Xu, Zhang \& Han (2005).

The best sample for the $\Sigma-D$ analysis consists of compact
SNRs from starburst galaxy M82 (see Arbutina et al. 2004 and Paper
I). The analyzed sample (21 SNRs) was taken from Huang et
al. (1994) and McDonald et al. (2002). This sample is
different than other Galactic and extragalactic samples because
it has the steepest $\Sigma-D$ slope (see Paper I) and shows a
relatively high degree of $L-D$ correlation (for more on $L-D$
correlation and trivial $\Sigma-D$ relation concept, see Arbutina
et al. 2004). Furthermore, unlike other samples, it consists
of a relatively high number of very small and very bright SNRs
(Paper I, Fenech et al. 2008, hereafter F08). It is important to
note that in an extragalactic sample all the SNRs are
essentially at the same distance, wherefore a uniformly sensitive
survey has a uniform  sensitivity in luminosity, or surface
brightness for all SNRs within the sample. The survey sensitivity
selection effect has weaker influence on the M82 $\Sigma-D$ slope
than on the slopes derived for other nearby galaxies because M82
SNRs are of relatively high brightness. This is shown in Paper I
on Monte Carlo generated artificial extragalactic samples; after
generating the sample,  a sensitivity cutoff is applied selecting
only the points above the survey sensitivity line. The apparent
(after selection) and true (before selection) fitted slopes of the
simulated samples are then compared with the slopes fitted to
the real data.

In this paper we use the new observations of compact SNRs in M82 by F08. From F08, we extract data set that consists of 31 SNRs and fit $\Sigma-D$ regression for both orthogonal and vertical offsets. Additionally, {Monte Carlo simulations are performed to illustrate the selection effect of survey sensitivity on} the fitted $\Sigma-D$ slopes for M82 sample (the simulation algorithm is explained in more details in Section 5).

\section{The M82 data sample}

 All data analyzed herein are collected from F08. The central kpc of M82 was mapped at 5 GHZ using MERLIN\footnote{Multi-Element Radio Linked Interferometer Network}. The largest detectable angular size with this array at 5 GHz is $\sim1.2$ arc seconds (18.5 pc at the distance of M82). Fenech et al. present new MERLIN observation made in 2002 along with observations made ten years earlier which previously published by Muxlow et al 1994. Depending on particular image parameters, the angular resolution in 1992 and 2002 data varies in the range $35-50$ mas ($0.54-0.78$ pc), but all sources were resolved with 35 mas beam (F08). Also, the rms noise in the 2002 data images varies in the range $17-24$ $\mathrm{\mu Jy beam^{-1}}$ and $46-60$ $\mathrm{\mu Jy beam^{-1}}$ for the 1992 data images (F08).

  The sources from 2002 observation are listed in Table 2 of F08. Out of 55 sources there are 36 SNRs. For the purpose of this paper, we have excluded five SNRs with the largest angular size diameter estimates (sources with peak flux $\lesssim 0.1~\mathrm{mJy~beam}^{-1}$). Inspection of Figure 3 in F08 shows that these five sources are mostly of non-compact structure with only the brightest parts above the sample sensitivity limit ($0.085~\mathrm{mJy~beam}^{-1}$). Consequently, these sources are easily confused with noise and the diameters of their faint extended structures can not be accurately estimated. This left us with the 31 SNR data set, referred to as S1 in further text. While all the data in S1 have associated integrated flux density errors, only 7 points have associated diameter errors. To calculate error in $\Sigma$ we need both flux density errors and diameter errors. Table 3 from F08 presents flux densities, diameters and associated errors for 10 SNRs observed with MERLIN in 1992 (not necessarily listed in Table 2 of F08). We used these 1992 measurements of flux densities, diameters and associated errors; this sample is further referred to as S2. For elliptical sources we calculated the mean geometric diameter (for both S1 and S2). The fits of non-weighted  vertical and orthogonal offsets for S1, and non-weighted and weighted vertical and orthogonal offsets for S2 are presented in the next Section. In Table 1 we present the range of relevant quantities for S1 and S2 samples.

\begin{table}
\caption{Summary characteristics of the selected samples (the number of sources in sample and range of relevant quantities). The adopted M82 distance is 3.2 Mpc (F08) which provides
a linear size equivalent to 1 mas $\equiv$ 0.0155 pc (as used in F08).}
\begin{tabular}{c|cc}
&S1&S2\\\hline
$N^a$&31&10\\
$S^b~[\mathrm{mJy}]$&$[0.099,19.435]$ &$[0.72,39.01]$\\
$L~[\mathrm{Js^{-1}{Hz}^{-1}}]$&$[1.21\times10^{17}, 2.38\times10^{19}]$&$[8.82\times10^{17} , 4.78\times10^{19}]$\\
$D~[\mathrm{pc}]$&$[0.64,6.2]$ &$[0.205,2.25]$\\
$\Sigma~[\mathrm{Wm^{-2}{Hz}^{-1}sr^{-1}}]$&$[9.24\times10^{-19},4.16\times10^{-15}]~$&$(6.42\times10^{-17},1.22\times10^{-13})$\\\hline
\end{tabular}\\
{\small Notes: $^a$ The number of sources in sample; $^b$ Integrated flux density.}
\end{table}

\section{Fitting}

  The standard fitting procedure in the $\Sigma-D$ plane based on the vertical (parallel to y-axis) $\chi^2$ regression has been used for calibration of empirical $\Sigma-D$ relations. Recently, Bandiera \& Petruk (2010) have used a different method -- regression analysis with two independent variables -- diameter $D$, and the density of environment $n_0$. In this paper, we stay with one independent variable $D$ (or $\Sigma$), but change fitting procedure from vertical to orthogonal offsets. Dependence on $n_0$ is important for $\Sigma-D$ analysis. M82 SNRs evolve in denser environment (Chevalier \& Fransson 2001, Arbutina \& Uro{\v s}evi{\' c} 2005), but variation in the ambient density certainly exists. This is consistent with observations of structural evolution and wide range in expansion velocities of individual SNRs in M82 (e.g. F08, Pedlar et al. 1999, McDonald et al. 2001, Beswick et al. 2006 and Fenech et al. 2010). Variation in expansion velocities is probably constrained by the differences in ambient density. This variation in density, if it is assumed that the SNRs are evolving along similar evolutionary tracks, probably does provide one of the key reasons for the moderately large scatter in the plotted $\Sigma-D$ correlation.

   For description of the radio surface brightness evolution of an SNR, we should investigate $\Sigma-D$ correlation, while for the distance determination of SNRs we need $D-\Sigma$ correlation (see Green 2009). The starting point of our analysis is the requirement that the $D-\Sigma$ and $\Sigma-D$ fit parameter values are invariant within the estimated uncertainties. This can be {achieved with} the orthogonal regression fitting procedure. Here, we use both types of fitting: standard (vertical) and orthogonal, and compare {the} results.

  Data fitting is performed numerically. We seek for the minimum of the $\chi^2$ function using the simplex algorithm (O'Neil 1971).  The fit  parameter values and their errors, presented in Tables 2-4, are the mean values and associated standard deviations after 10000 bootstrap data re-samplings for each fit. When fitting with data errors, the vertical offsets are  weighted with $\sigma_{y_i}^2$, while orthogonal offsets $\chi^2$ are calculated as  $\frac{(y_i-A-\beta x_i)^2}{\sigma_{y_i}^2+(\sigma_{x_i}\beta)^2}$.

\section{Analysis of fit statistics}

 At first glance, inspection of Figures 1 and 2 and Tables 2-4 leads to the conclusion that resulting fit parameters values are significantly
 influenced with the type of the fitting procedure. The $\Sigma-D$ slopes are obviously steeper for orthogonal offsets (Figures 1 and 2; Table 2).
 The approximately "trivial" $\Sigma-D$ slope $(\beta\approx2.4)$ is transformed into a very steep slope $\approx3.9$ for the S1 sample of 31 SNRs
 (Table 2). Also, for the poorer sample with respect to the number of objects (10 S2 SNRs), steeper slopes are obtained, but the differences are not
 so huge as in the case of the larger sample (see Table 2). On the other hand, $D-\Sigma$ slopes are approximately the same in both fitting procedures
 (see Table 4). This is due to a rather small span of diameters (one order of magnitude) in comparison to the span of surface brightnesses
 (four orders of magnitude). This leads to flatter slopes which results in similar lengths of vertical and orthogonal offsets giving similar fit parameters.

For the proper $\Sigma-D$ analysis, the $L-D$ correlation should be checked. If the $L-D$ correlation does not exist, the trivial
 $\Sigma\propto D^{-2}$ form should not be used (Arbutina et al. 2004).
 The statistics of $L-D$ correlations for both M82 samples are rather poor. For the S2 sample , this is because of the relatively low coefficient of correlation and a small number of objects in the sample, while for the S1 sample, because of a very low coefficient of correlation. The coefficient of correlation $r$ is calculated using the following equation:

\begin{equation}
 r = \frac{\sum_i(x_i-\overline{x})(y_i-\overline{y})}{\sqrt{\sum_i (x_i-\overline{x})^2}\sqrt{\sum_i(y_i-\overline{y})^2}}.
\end{equation}

\noindent Finally, based on the poor statistical
results of the $L-D$ fits (Table 3) it can be concluded that
both extracted samples show a high degree of scattering.

\section{Monte Carlo simulations}

We performed a set of Monte Carlo simulations to estimate the
influence of the survey sensitivity selection effect on the
$\Sigma-D$ slope for M82 sample (31 SNRs). In both fitting
procedures (vertical and orthogonal) we used the algorithm
described below.

\subsection{Vertical offsets}

Monte Carlo simulations are similar to those described in Paper I.
First, we determined the empirical $\log \Sigma$ standard
deviation from the best fit line, assuming $\log D$ as the
independent variable. We then selected an interval in $\log D$
between 0.65 -- 100 pc. This interval is then sprinkled with
random points of the same $\log D$ density as that of the real
data.

The simulated points, that lie on the $\log D$ axis, are then projected
onto a series of lines at different slopes (in steps of 0.1
from 1.5 to 4.5). Each of these lines passes through the extreme
upper left hand end of the best fit line to the real data. We
also added Gaussian noise in $\log \Sigma$, which is related to the scatter
of the real data by a parameter called "scatter". A scatter of
1 corresponds to the same standard deviation as that of the real
data.

An appropriate sensitivity cutoff is applied to the simulated data
points, selecting points above the sensitivity line (for
simplicity, we assumed a sensitivity line that passes through the
real data point of the lowest brightness). This is done 1000
times for each simulated slope and a least squares best fit line
(vertical regression) is generated for artificial samples.

In Table 5, the first column lists the scatter and second column shows the value of the simulated
slope. Columns 3-6 are for vertical offsets, the mean and standard deviation of the best fit
slopes for the generated samples and mean and standard deviation of the best fit slopes for sensitivity
selected generated samples, respectively. In the same manner, columns 7-10 list the properties for orthogonal offsets. Figure 3 shows one of our Monte Carlo generated samples for vertical offsets at 5 GHz with a scatter of 1 and the simulated slope of 2.4.


\subsection{Orthogonal offsets}

 We calculated standard deviation of data from the data
best fit line using the orthogonal offsets. Then we have
generated random diameters as described above. The
points are then projected onto the simulated slope line.
Then we added the Gaussian noise
to the simulated points in orthogonal direction from the simulated slope line,  as:

\begin{equation}
D_\mathrm{noise} = D_\mathrm{proj} \pm \frac{n_{\rm orth.}}{\sqrt{1+\frac{1}{b^2}}},\quad
\Sigma_\mathrm{noise} = \Sigma_\mathrm{proj} \pm \frac{n_{\rm orth.}}{1+b^2},
\end{equation}

\noindent with $n_{\rm orth.}$ being the noise in the orthogonal
offset direction and $b$ the simulated slope. All artificial
samples are fitted using orthogonal fitting procedure
repeated 1000 times. The results of Monte Carlo simulations for
orthogonal offsets are presented in Figure 3 (for slope
$\beta=3.9$) and Table 5.

\section{Discussion}

By contrast to the standard (vertical) fitting, the
orthogonal regression procedure leads to a significant change in
the slope of the $\Sigma-D$ relation from 2.4 to 3.9 (Table 2).
The latter is a steep empirical slope, very far away from
the trivial one ($\beta\approx2$), and between theoretical
predictions for the energy conserving phase of an SNR evolution
($\beta=3.5$ and $\beta=4.25$), obtained by Duric \& Seaquist
(1986) and Berezhko \& V{\" o}lk (2004), respectively. {The
inverted slope value} (1/3.9 $\approx$ 0.26) is approximately the
same as {the} value obtained by $D-\Sigma$ fitting. {Thus, for the
orthogonal fitting procedure, $D-\Sigma$ and $\Sigma-D$ fit
parameters values} are invariant within the estimated
uncertainties (see Tables 2, 4). A careful inspection of Table 4
leads to the conclusion that both fitting procedures provide
similar $D-\Sigma$ slopes. Therefore, in case of M82 SNRs,
instead of using the more complicated orthogonal regression
method, one can find $D-\Sigma$ slope by the standard
(vertical) fitting and after that invert it to find a valid
$\Sigma-D$ slope. This supports a suggestion to use
$D-\Sigma$ relation} given by Green (2009). This is possible
because of the narrow span of diameters for M82 SNRs (one order of
magnitude) in comparison with the wide span of brightnesses (four
orders of magnitude). If these spans are similar, the orthogonal
procedure has to be used for the useful $D-\Sigma$ regression,
too.

Based on the $L-D$ analysis (Table 3) it can be concluded
that corresponding correlations are very poor. A large
scatter in the data is evident and hence the correlation
coefficients are low. On the other hand, some poor trends in the
$LD$ plane are visible but the moderately large level of
scattering (or a small number of objects) in analyzed
samples could not provide any valid conclusion about these trends
(see Figures 1 and 2, second panel). There are observed
differences in the expansion velocities of the SNRs in M82 (see
F08 and references therein). This implies either that they are on
different evolutionary tracks (connected with different initial
energy of explosion), and/or expanding into different density
regions, or as a consequence may be in different phases of SNR
evolution. Consequently, a relatively large data scatter can be
explained by the above noted influences.

When presenting fit parameter values in the tables, we have given
the ratio of weighted sum of square residuals (offsets) and number
of degrees of freedom ($WSSR/ndof$). The probability $Q$ of
obtaining larger weighted sum of square residuals is also
presented. While for non-weighted offsets $WSSR/ndof$ and
$Q$ values are of no practical importance and are calculated only
for the sake of completeness, they show that weighted fits are not
statistically justified ($Q\gtrsim0.001$ and $WSSR/ndof \sim 1$,
when the scatter is of the order of $\sim1$ standard deviation).


 Arbutina \& Uro\v sevi\'c (2005) argued that SNRs of different types can be found along more or less parallel tracks in the $\Sigma D$ plane. The tracks are presumably defined by the density of the surrounding environment in which SNRs evolve.
 Inspection of Table 5 shows that for scatters larger than 1, slopes of the $\Sigma-D$ relation are seriously under the influence of the sensitivity cutoff. This implies that for a reliable calibration of the $\Sigma-D$ relation, compact samples should be used (SNRs with similar initial properties evolving in similar environments). This criteria is probably satisfied for the M82 SNR sample consisting of young SNRs that evolve in dense environment of M82 starburst region. The latter conclusions may be valid if all SNRs have entered the energy conserving (Sedov) phase. The exact phase of evolution remains the main uncertainty for M82 SNRs. At least one compact SNR 43.3+59.2 has an exponent $m$, from the dynamical law $R\propto t^m$, $\ge 0.68$ implying the free expansion (Beswick et al. 2006). Chevalier \& Fransson (2001), on the other hand, argue that M82 SNRs may even be in the radiative phase.

  For the simulated data scatter of 1, that should resemble the real scatter of the data, the slope of the $\Sigma-D$ relation is not severely biased by the sensitivity cutoff. Similar conclusion is drawn from the Monte Carlo sensitivity related simulations in Paper I. They used the M82 data sample of Huang et al. (1994), collected with the Very Large Array, while the M82 sample analyzed in this work was recorded with the MERLIN measurements. This resulted in somewhat different sensitivity lines but nevertheless both studies came up with similar conclusions.

Monte Carlo simulations are carried out for the purpose
of checking the completeness of the M82 SNR sample. Objects
with low surface brightnesses can not be detected because they are
affected by the survey sensitivity selection effect. By simulating
this effect, we tried to find out whether our sample (i.e. corresponding
$\Sigma-D$ slope) is representative for M82 SNR population or not.
Inspection of Table 5, when scatter is generated by
vertical offsets, show that sensitivity selection effect makes the
observed $\Sigma-D$ slopes shallower. The result is
identical to the one obtained in Paper I. When scatter is
generated by the orthogonal offsets, the sensitivity line does not
cut a significant number of artificial objects located in
lower-left part of the field. In scatter 1 scenario, the
sensitivity cutoff does not affect $\Sigma-D$ slope (see Table
5). A very interesting situation arises in the simulation
of the orthogonal scatter 2 scenario. The $\Sigma-D$ slopes are
changed significantly. The lower-left part of artificial samples
is cut by the sensitivity line when scatter is high (higher than
real one) and the slopes of relations become shallower. Based
on the analysis of the results of simulations presented (Table 5),
we believe that the orthogonal scatter 1 scenario is more likely
for two reasons: (1)  the slope ($\beta=3.9$) is obtained by the
orthogonal procedure that gives the invariant $\Sigma-D$ and
$D-\Sigma$ slopes, and (2) scatter is generated by the orthogonal
offsets and corresponds to the real scatter in observed
data-set. We conclude that sensitivity selection effect does
not {have a major impact on} the $\Sigma-D$ {slope} for M82 SNRs.

With $D-\Sigma$ and $\Sigma-D$ fit slopes being invariant within
the estimated uncertainties in the orthogonal fitting procedure,
assuming a relatively complete sample, the $\Sigma\propto
D^{-3.9}$ relation for M82 SNRs could potentially describe the
evolution of young SNRs in the energy conserving phase of their
evolution, and this relation might be useful for estimating
distances to such SNRs. The problem that remains is the coupling
of the evident data scattering in F08 SNR sample and a small
number of objects for which reliable statistics can be done.
Another problem is that most of the sources do not show
significant flux density variation (Kromberg et al. 2000),
implying trivial physical relation $L_\nu\approx$ const. Some
sources, like 41.30+59.6 show flux increase, rather than decrease
(F08). Therefore the $\Sigma-D$ relations obtained in this paper
should be used with caution.

Finally, some compact radio objects in M82 may not be SNRs,
as proposed by Seaquist \& Stankovi{\'c} (2007). They analyzed
compact nonthermal radio objects and concluded that some of them
are probably the so-called Wind Driven Bubbles (WDBs) due to the
lack of observed time variability in most of the sources, implying
ages greater than expected for SNRs. However, the {recent}
detection of $\gamma$ radiation from M82 (Abdo et al. 2010)
confirms standard opinion that radio objects in M82 are indeed
SNRs. The strong shock waves of young SNRs are necessary for the
efficient production of cosmic rays by the so-called diffuse shock
acceleration (DSA) mechanism. The inverse Compton scattering of
the background electromagnetic radiation by the cosmic ray
electrons (leptonic model) or a decay of neutral pions, mainly
produced by cosmic ray protons during the interaction with the gas
(hadronic model) represent two basic mechanisms for production of
$\gamma$ rays. WDBs probably do not represent proper sites for the
production of $\gamma$ rays, due to slower shock waves in
comparison to shock waves of young SNRs.

\section{Conclusions}

 We suggest the orthogonal regression procedure to be used for
obtaining empirical $\Sigma-D$ relations. In that case the values
of parameters obtained from fitting of $\Sigma-D$ and $D-\Sigma$
relations are invariant within estimated uncertainties.
Alternatively, if a data span in $\Sigma$ covers more
orders of magnitude than a data span in $D$, fitting of the
$D-\Sigma$ relation with vertical offsets can give $\beta$ that
resemble the slope fitted with either $\Sigma-D$ or $D-\Sigma$
orthogonal offsets. The steep $\Sigma-D$ slope $(\beta=3.9)$ is
obtained {when fitting the orthogonal regression to the updated
M82 SNR sample.} The results of our Monte Carlo simulations
suggest that this slope is probably free of the sensitivity
selection effect. Moreover, it is closer to the updated
theoretically derived slopes for the energy conserving phase of
SNR evolution. The relation $\Sigma\propto D^{-3.9}$ could
represent the average evolutionary track for SNRs in M82, and
could potentially be used for estimating the distances of
young SNRs expanding in dense environment. However, data
scattering and, more importantly, a relatively small number of
objects in the analyzed samples constrain the reliability of this
relation. Due to this, the obtained $\Sigma-D$ relations should
be used with caution. More observations and better theoretical
description are necessary for deeper understanding of the
radio evolution of these SNRs.

\acknowledgments {The authors would like to thank Dragana
Momi\'{c} for reading the manuscript and the anonymous referee for
valuable comments that improved the quality of this paper. This
work is part of the Projects No. 146003 and 146012 supported by
the Ministry of Science and Environmental Protection of Serbia.\\\\

{\large \textbf {References}}\\

 Abdo et al. 2010, ApJl,709, L152

 Arbutina, B., Uro{\v s}evi{\' c}, D., Stankovi{\' c}, M.
\& Te{\v s}i{\' c}, Lj. 2004, MNRAS, 350, 346

 Arbutina, B. \& Uro{\v s}evi{\' c}, D. 2005, MNRAS, 360, 76

 Bandiera, R.  \& Petruk O. 2010, A \& A, 509, A34

 Berezhko, E.G. \& V{\" o}lk, H.J. 2004, A \& A, 427, 525

 Beswick, R. J. et al. 2006, MNRAS, 369, 1221

 Case, G. L. \& Bhattacharya, D. 1998, ApJ, 504, 761

 Chevalier, R. A. \& Fransson, C. 2001, ApJl, 558, L27

 Chomiuk, L. \& Wilcots, E.M. 2009, \aj, 137, 3869

 Clark, D.H. \& Caswell, J.L. 1976, MNRAS, 174, 267

 Duric, N. \& Seaquist, E.R. 1986, ApJ, 301, 308

 Fenech, D.M., Muxlow, T.W.B., Beswick, R.J., Pedlar, A. \& Argo, M.K. 2008, MNRAS, 391, 1384 (F08)

 Fenech, D., Beswick, R., Muxlow, T. W. B., Pedlar, A. \& Argo, M. K. 2010, astro-ph/1006.1504

 Green, D. A. 1984, MNRAS, 209, 449

 Green, D.A. 2009, Bull. Astr. Soc. India, 37, 45

 Guseinov, O.H., Ankay, A., Sezer, A. \& Tagieva, S.O. 2003, Astron. Astrophys. Transactions, 22, 273

 Huang Z. P., Thuan T. X., Chevalier R. A., Condon J. J. \& Yin Q. F. 1994,
ApJ, 424, 114

 Kronberg P. P., Sramek R. A., Birk G. T., Dufton Q. W., Clarke T. E. \& Allen
M. L. 2000, ApJ, 535, 706

 McDonald, A. R., Muxlow, T. W. B., Pedlar, A., Garrett, M. A., Wills, K. A., Garrington, S. T., Diamond, P. J. \& Wilkinson, P. N. 2001, MNRAS, 322, 100

 McDonald, A. R., Muxlow, T. W. B., Wills, K. A., Pedlar, A. \&
Beswick, R. J. 2002, MNRAS, 334, 912

 Milne, D.K. 1970, Austral. J. Phys. 23, 425

 Milne, D.K. 1979, Austral. J. Phys. 32, 83

 Muxlow, T. W. B., Pedlar, A., Wilkinson, P. N., Axon, D. J., Sanders, E. M. \& de Bruyn, A. G. 1994, MNRAS, 266, 455

 O'Neil R. 1971, Journal of the Royal Statistical Society Series C (Applied Statistics), 20(3), 338-345

 Poveda, A. \& Woltjer, L. 1968, AJ, 73(2), 65

 Pedlar, A. Muxlow, T. W. B., Garrett, M. A., Diamond, P., Wills, K. A., Wilkinson, P. N. \& Alef, W. 1999, MNRAS, 307, 761

 Seaquist, E.R. \& Stankovi{\'c}, M. 2007, ApJ, 659, 347

 Uro{\v s}evi{\' c}, D. 2002, Serb. Astron. J., 165, 27

 Uro{\v s}evi{\' c}, D., Pannuti, T. G., Duric, N. \&  Theodorou,
A. 2005, A \& A, 435, 437 (Paper I)

 Xu, J.-W., Zhang, X.-Z. \& Han, J.-L. 2005, Chin. J. Astron. Astrophys., 5, 165


\begin{center}
\begin{figure}
\includegraphics[width=0.5\textwidth]{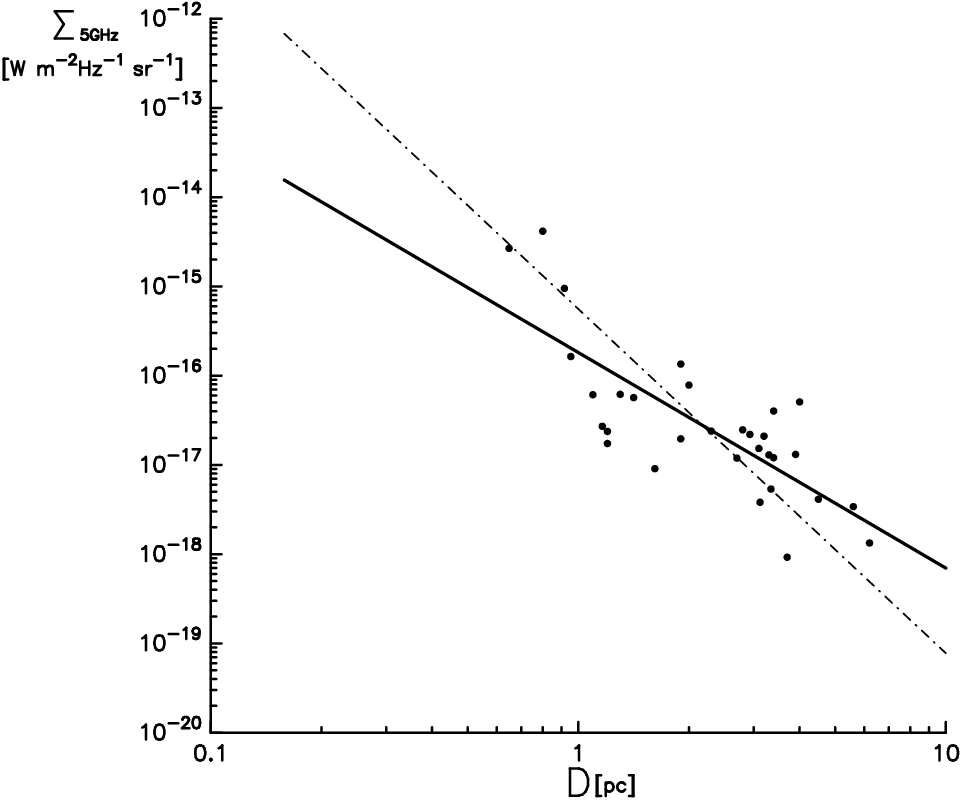}
\includegraphics[width=0.5\textwidth]{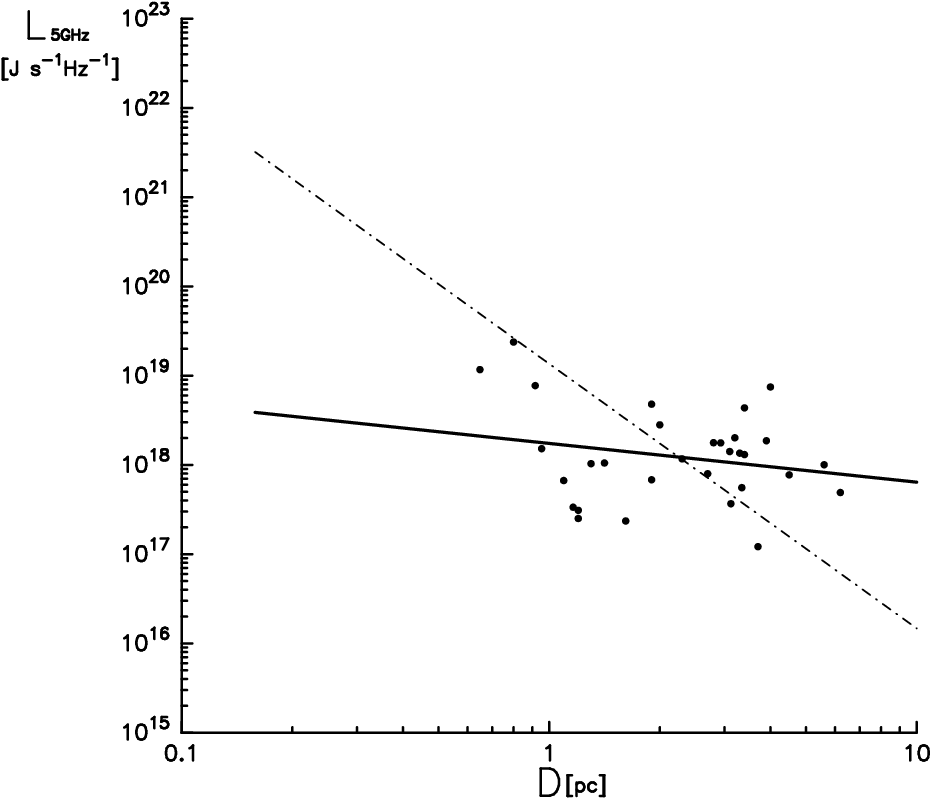}
\vspace*{0.5in}
\hspace*{0.30in}\includegraphics[width=0.5\textwidth]{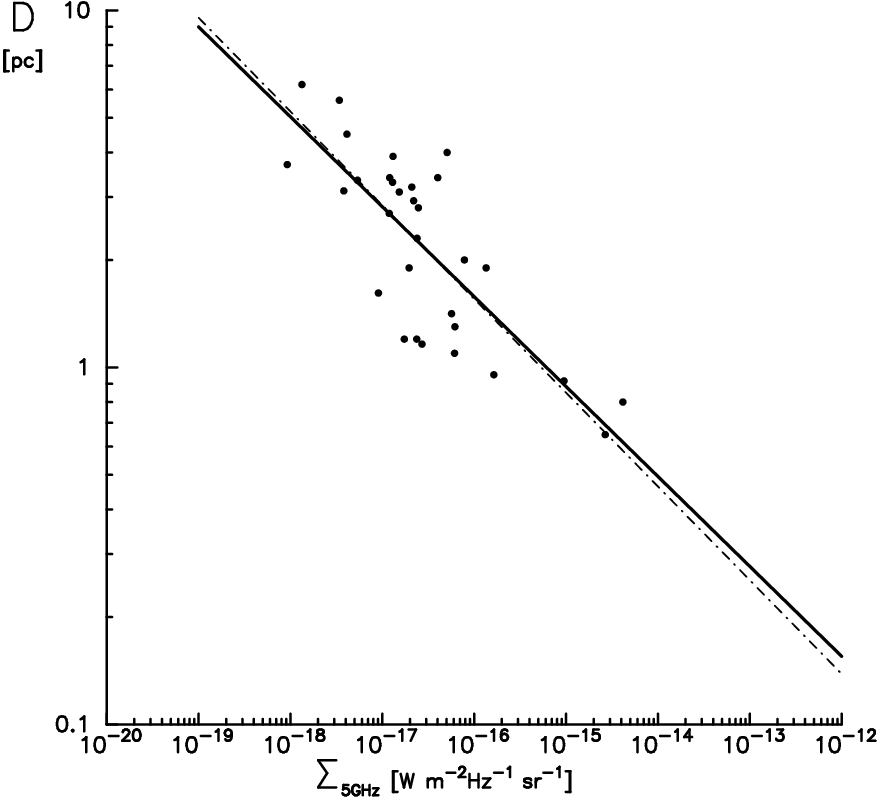}
\caption{Data from Table 2 in F08 (31 SNRs). Thick solid line -- non-weighted vertical offsets; dashed-doted line -- non-weighted orthogonal offsets.}
\end{figure}

\begin{figure}
\includegraphics[width=0.5\textwidth]{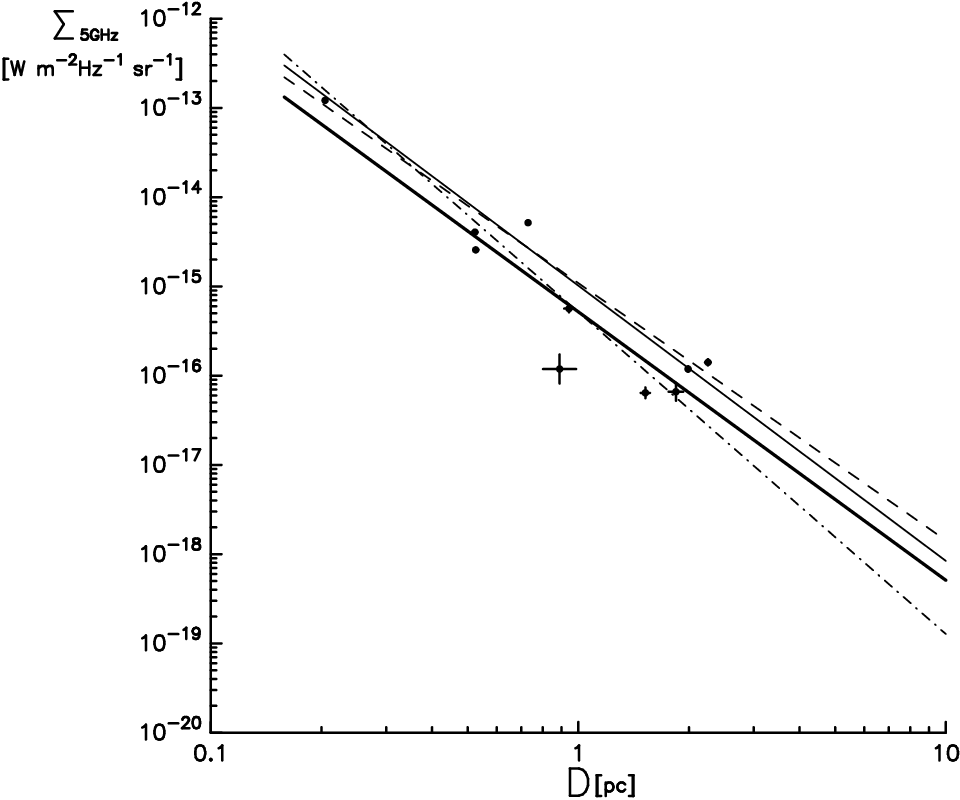}
\includegraphics[width=0.5\textwidth]{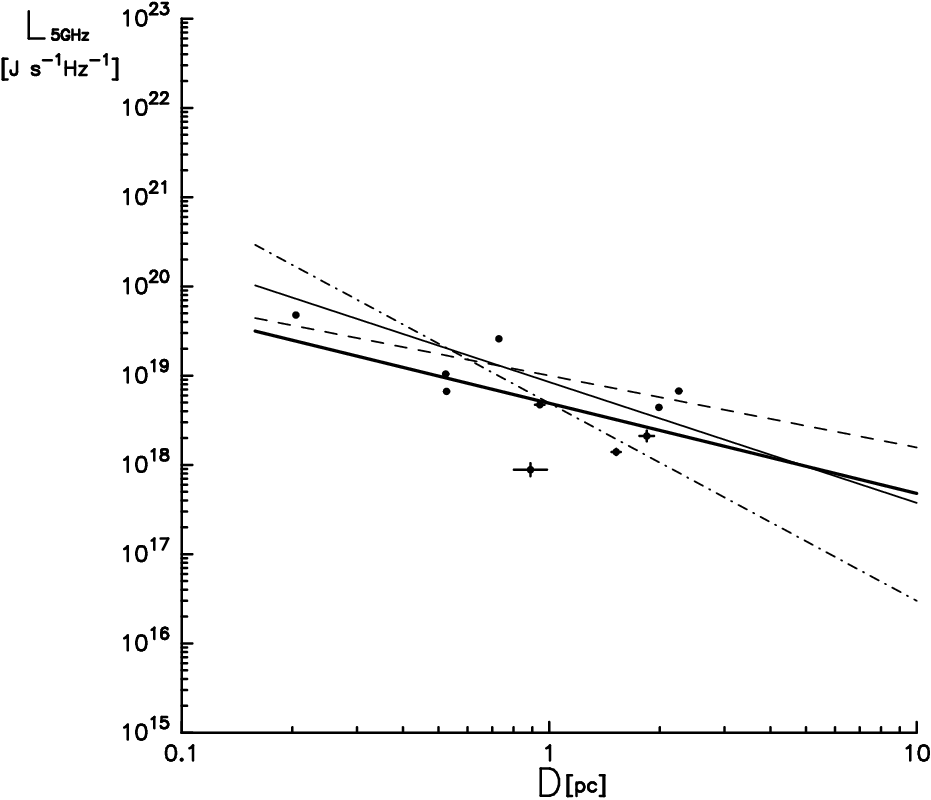}\\
\vspace*{0.5in}
\hspace*{0.30in}\includegraphics[width=0.5\textwidth]{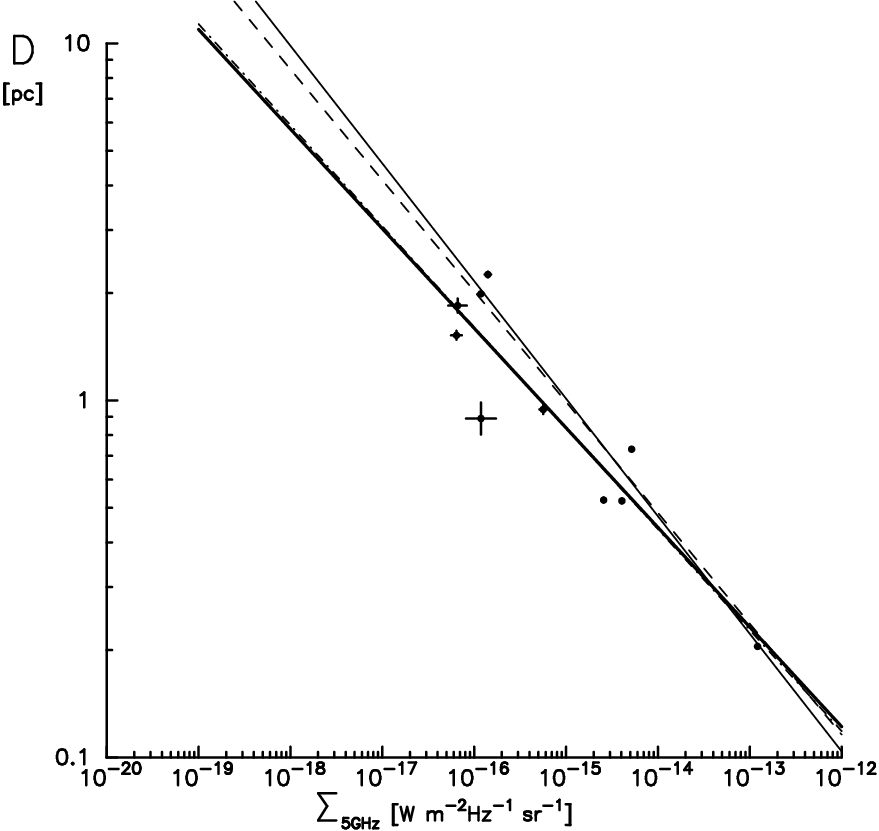}
\caption{Data from Table 3 in F08 (10 SNRs). The errors are plotted for all points. Thick solid line -- non-weighted vertical offsets; dashed line -- weighted vertical offsets; dashed-doted line and thin solid line are for the non-weighted and weighted orthogonal offsets, respectively.}
\end{figure}
\end{center}

\begin{figure}
\includegraphics[width=0.5\textwidth]{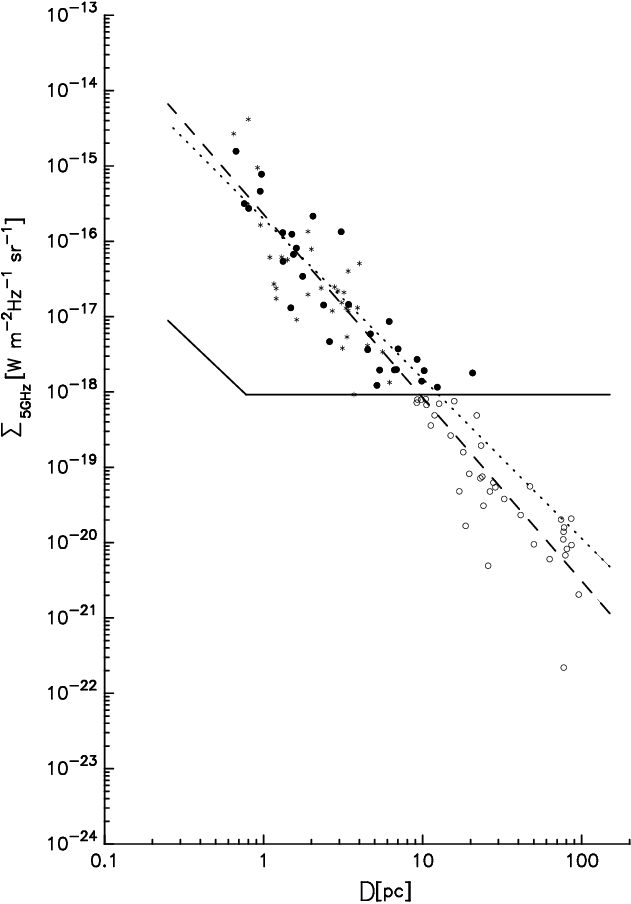}
\includegraphics[width=0.5\textwidth]{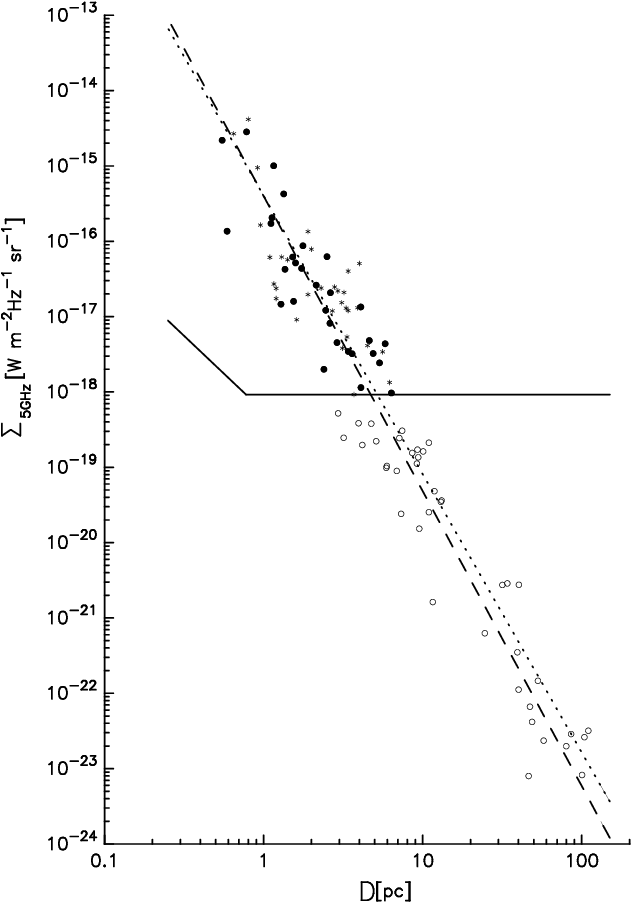}\\
\caption{The Monte Carlo generated sample at 5 GHz for a scatter
of 1. M82 data points (31 SNRs, signed by asterisks) are plotted
together with the sensitivity (solid) line; artificially generated
points are plotted above (filled circles) and below (open circles)
this line. Dashed line -- fit before selection; doted line -- fit
after selection. Left -- vertical offsets for a simulated slope of
$2.4$; right -- orthogonal offsets for a simulated slope of $3.9$.
}
\end{figure}

\begin{table}\small
\caption{The $\Sigma-D$ relation.}
\begin{tabular}{l|ccccccc}
\hfil fit \hfil               &  $\log A$           &        $\Delta\log A$    &       $\beta$  &    $\Delta\beta$    &     $Q$       &   $WSSR/ndof$ & $\sqrt{WSSR/ndof}$\\\hline\hline
\multicolumn{8}{c}{The sample of 10 SNRs  from Table 3 in F08, $r = -0.924164,  ~ r^2 = 85.407974\%$.}\\\hline
Ver. os.       & -15.2842 & 1.48035e-01  & 3.00753 &  0.43692 & 9.92070e-01 & 1.92172e-01 & 4.38374e-01 \\
Ver. os. w. & -14.9625 & 1.99015e-01  & 2.88255 &  0.41673 & 0.00000e+00 & 2.08336e+02 & 1.44338e+01 \\
Ort. os.       & -15.2874 & 1.56158e-01  & 3.60747 &  0.52914 & 9.99999e-01 & 1.62385e-02 & 1.27430e-01 \\
Ort. os. w. & -14.9921 & 1.82236e-01  & 3.08412 &  0.47155 & 1.07173e-201 & 1.19871e+02 & 1.09485e+01 \\
\hline
\multicolumn{8}{c}{The sample of 31 SNRs from Table 2 in F08, $r = -0.782763,  ~ r^2 = 61.271836\%$.}\\\hline
Ver. os.        & -15.7409 & 2.01064e-01  & 2.41576 &  0.43166 & 9.99967e-01 & 2.71169e-01 & 5.20739e-01 \\
Ort. os.        & -15.2535 & 1.87001e-01  & 3.85631 &  0.43339 & 1.00000e+00 & 2.58510e-02 & 1.60783e-01 \\
\hline
\end{tabular}
 \end{table}

\begin{table}\small
\caption{The $L=BD^{-\delta}$ relation.}
\begin{tabular}{l|ccccccc}
\hfil fit \hfil               &  $\log B$           &        $\Delta\log B$    &       $\delta$  &    $\Delta\delta$    &     $Q$       &   $WSSR/ndof$ & $\sqrt{WSSR/ndof}$\\\hline\hline
\multicolumn{8}{c}{The sample of 10 SNRs from Table 3 in F08, $r = -0.641483,  ~ r^2 = 41.150074\%$}\\\hline
Ver. os.       & 18.6909 & 1.40852e-01  & 1.01038 &  0.43436 & 9.92093e-01 & 1.92008e-01 & 4.38187e-01 \\
Ver. os. w. & 19.0008 & 2.31966e-01  & 0.80511 &  0.57985 & 0.00000e+00 & 1.62263e+03 & 4.02818e+01 \\
Ort. os.       & 18.6929 & 3.11494e-01  & 2.21533 &  0.95097 & 9.99888e-01 & 5.97075e-02 & 2.44351e-01 \\
Ort. os. w. & 18.9292 & 2.20137e-01  & 1.35415 &  0.82877 & 0.00000e+00 & 8.40500e+02 & 2.89914e+01 \\
\hline
\multicolumn{8}{c}{The sample of 31 SNRs from Table 2 in F08, $r = -0.236428,  ~ r^2 = 5.589839\%$}\\\hline
Ver. offst.       & 18.2409 & 2.01469e-01  & 0.43333 &  0.43048 & 9.99971e-01 & 2.68603e-01 & 5.18269e-01 \\
Ort. offst.       & 19.1334 & 7.52688e-01  & 2.96520 &  1.40621 & 1.00000e+00 & 7.25849e-02 & 2.69416e-01 \\
\hline
\end{tabular}
 \end{table}

 \begin{table}\small
\caption{The $D-\Sigma$ relation.}
\begin{tabular}{l|ccccccc}
\hfil fit \hfil               &  coef.           &        $\Delta{\rm coef.}$    &       $1/\beta$  &    $\Delta(1/\beta)$    &     $Q$       &   $WSSR/ndof$ & $\sqrt{WSSR/ndof}$\\\hline\hline
\multicolumn{8}{c}{The sample of 10 SNRs from Table 3 in F08, $r = -0.924164,  ~ r^2 = 85.407974\%$}\\\hline
Ver. os.       & -4.26141 & 5.54597e-01  & 0.27896 &  0.03736 & 9.99999e-01 & 1.74798e-02 & 1.32211e-01 \\
Ver. os. w. & -4.66426e & 6.73978e-01  & 0.31067 &  0.04486 & 0.00000e+00 & 2.80880e+02 & 1.67595e+01 \\
Ort. os.       & -4.32109e & 5.48758e-01  & 0.28290 &  0.03697 & 9.99999e-01 & 1.61987e-02 & 1.27274e-01 \\
Ort. os. w. & -4.93663 & 6.22588e-01  & 0.32945 &  0.04178 & 2.32472e-200 & 1.19097e+02 & 1.09131e+01 \\
\hline
\multicolumn{8}{c}{The sample of 31 SNRs from Table 2 in F08, $r = -0.782763,  ~ r^2 = 61.271836\%$.}\\\hline
Ver. os.       & -3.83090 & 4.62411e-01  & 0.25182 &  0.02793 & 1.00000e+00 & 2.75240e-02 & 1.65904e-01 \\
Ort. os.       & -4.00880 & 5.02672e-01  & 0.26253 &  0.03021 & 1.00000e+00 & 2.58590e-02 & 1.60807e-01 \\
\hline
\end{tabular}
 \end{table}

\newpage

\appendix

\section{On-line material: Results of Monte Carlo simulations}

\begin{table}
\scriptsize
\caption{The results of Monte Carlo simulations}
\begin{tabular}{p{1.0cm}p{1.4cm}p{1.4cm}p{1.4cm}p{1.4cm}p{1.4cm}p{1.4cm}p{1.4cm}p{1.4cm}p{1.4cm}}\hline
&&\multicolumn{4}{|c|}{Vertical offsets}&\multicolumn{4}{|c}{Orthogonal offsets}\\\hline
Scatter & The slope that is simulated & Mean simulated slope & Standard deviation of mean simulated slope & Mean slope after selection & Standard deviation of slope after selection & Mean simulated slope & Standard deviation of mean simulated slope & Mean slope after selection & Standard deviation of slope after selection\\
\hline
1.0	&	1.500000	&	1.500696	&	0.099013	&	1.286365	&	0.109226	&	1.497463	&	0.055816	&	1.497168	&	0.055705	\\
1.0	&	1.600000	&	1.600147	&	0.104393	&	1.349833	&	0.123326	&	1.598297	&	0.057974	&	1.587014	&	0.058188	\\
1.0	&	1.700000	&	1.699489	&	0.099988	&	1.426584	&	0.131765	&	1.702749	&	0.060101	&	1.683068	&	0.067908	\\
1.0	&	1.800000	&	1.802722	&	0.101126	&	1.507596	&	0.148833	&	1.804823	&	0.063576	&	1.782269	&	0.076331	\\
1.0	&	1.900000	&	1.899529	&	0.101864	&	1.576387	&	0.156577	&	1.901645	&	0.065447	&	1.878356	&	0.082454	\\
1.0	&	2.000000	&	1.997602	&	0.096299	&	1.672339	&	0.168350	&	2.003504	&	0.066718	&	1.979787	&	0.094530	\\
1.0	&	2.100000	&	2.101995	&	0.101180	&	1.751897	&	0.183618	&	2.097585	&	0.069813	&	2.071091	&	0.112198	\\
1.0	&	2.200000	&	2.203092	&	0.101992	&	1.836862	&	0.198541	&	2.201668	&	0.070944	&	2.182622	&	0.121677	\\
1.0	&	2.300000	&	2.298177	&	0.100645	&	1.911342	&	0.212355	&	2.297719	&	0.078436	&	2.278483	&	0.139566	\\
1.0	&	2.400000	&	2.402304	&	0.099143	&	2.005975	&	0.228506	&	2.402684	&	0.080858	&	2.384728	&	0.146849	\\
1.0	&	2.500000	&	2.500295	&	0.100127	&	2.098402	&	0.243022	&	2.504435	&	0.081318	&	2.481287	&	0.165271	\\
1.0	&	2.600000	&	2.602096	&	0.101548	&	2.172323	&	0.261917	&	2.600812	&	0.088288	&	2.576799	&	0.184526	\\
1.0	&	2.700000	&	2.702694	&	0.101349	&	2.266196	&	0.287042	&	2.704265	&	0.085594	&	2.675869	&	0.204002	\\
1.0	&	2.800000	&	2.799537	&	0.100410	&	2.340753	&	0.298422	&	2.799311	&	0.090789	&	2.784884	&	0.229936	\\
1.0	&	2.900000	&	2.897140	&	0.098674	&	2.421775	&	0.310348	&	2.899872	&	0.095418	&	2.877078	&	0.238608	\\
1.0	&	3.000000	&	2.996749	&	0.099588	&	2.498980	&	0.323027	&	3.003264	&	0.098458	&	2.979326	&	0.249424	\\
1.0	&	3.100000	&	3.104522	&	0.098841	&	2.589717	&	0.345333	&	3.102771	&	0.102963	&	3.076613	&	0.290303	\\
1.0	&	3.200000	&	3.202747	&	0.102624	&	2.688157	&	0.357554	&	3.204156	&	0.104733	&	3.179113	&	0.326909	\\
1.0	&	3.300000	&	3.298963	&	0.099057	&	2.763613	&	0.362475	&	3.300337	&	0.106178	&	3.290091	&	0.310472	\\
1.0	&	3.400000	&	3.402224	&	0.102481	&	2.839851	&	0.412462	&	3.405014	&	0.104106	&	3.377635	&	0.405915	\\
1.0	&	3.500000	&	3.507588	&	0.104501	&	2.953550	&	0.443178	&	3.500932	&	0.107231	&	3.498262	&	0.398552	\\
1.0	&	3.600000	&	3.601357	&	0.098950	&	3.027741	&	0.440752	&	3.603510	&	0.113465	&	3.603639	&	0.441060	\\
1.0	&	3.700000	&	3.698463	&	0.102682	&	3.100663	&	0.431116	&	3.699302	&	0.115896	&	3.686992	&	0.517250	\\
1.0	&	3.800000	&	3.800653	&	0.102899	&	3.198014	&	0.492386	&	3.809688	&	0.118302	&	3.789941	&	0.506556	\\
1.0	&	3.900000	&	3.901953	&	0.100244	&	3.276876	&	0.484156	&	3.907993	&	0.121222	&	3.879406	&	0.624074	\\
1.0	&	4.000000	&	3.997578	&	0.100347	&	3.371390	&	0.524299	&	4.003947	&	0.128336	&	3.956310	&	0.633162	\\
1.0	&	4.100000	&	4.098606	&	0.103149	&	3.451102	&	0.567184	&	4.101723	&	0.127667	&	4.003518	&	0.836152	\\
1.0	&	4.200000	&	4.203367	&	0.099424	&	3.549554	&	0.561571	&	4.205925	&	0.131813	&	4.103016	&	0.918963	\\
1.0	&	4.300000	&	4.296356	&	0.103596	&	3.563594	&	0.564266	&	4.301982	&	0.133642	&	4.140517	&	0.947753	\\
1.0	&	4.400000	&	4.399762	&	0.102086	&	3.677029	&	0.617068	&	4.396678	&	0.136565	&	4.231672	&	1.092972	\\
1.0	&	4.500000	&	4.500762	&	0.099421	&	3.754998	&	0.640986	&	4.513678	&	0.138215	&	4.261958	&	1.206376	\\
2.0	&	1.500000	&	1.497074	&	0.201499	&	1.010993	&	0.208958	&	1.503948	&	0.123884	&	1.491782	&	0.120576	\\
2.0	&	1.600000	&	1.601748	&	0.206268	&	1.059087	&	0.211039	&	1.611962	&	0.125516	&	1.565151	&	0.122074	\\
2.0	&	1.700000	&	1.699103	&	0.206710	&	1.097170	&	0.225302	&	1.709634	&	0.127692	&	1.629440	&	0.129925	\\
2.0	&	1.800000	&	1.815947	&	0.198277	&	1.148707	&	0.227667	&	1.806221	&	0.136550	&	1.718790	&	0.151350	\\
2.0	&	1.900000	&	1.898634	&	0.205865	&	1.184780	&	0.240272	&	1.906057	&	0.140358	&	1.818100	&	0.177176	\\
2.0	&	2.000000	&	2.003181	&	0.208126	&	1.229978	&	0.239254	&	2.001608	&	0.141960	&	1.909523	&	0.193236	\\
2.0	&	2.100000	&	2.106723	&	0.196232	&	1.282816	&	0.266664	&	2.115680	&	0.149275	&	2.012858	&	0.243819	\\
2.0	&	2.200000	&	2.195067	&	0.200040	&	1.317523	&	0.278585	&	2.214054	&	0.154374	&	2.100802	&	0.277017	\\
2.0	&	2.300000	&	2.297552	&	0.198016	&	1.378041	&	0.304825	&	2.310805	&	0.166987	&	2.184658	&	0.346117	\\
2.0	&	2.400000	&	2.402734	&	0.192766	&	1.457052	&	0.326163	&	2.403606	&	0.169570	&	2.268059	&	0.405243	\\
2.0	&	2.500000	&	2.500341	&	0.191161	&	1.510620	&	0.345794	&	2.519952	&	0.178789	&	2.328069	&	0.539537	\\
2.0	&	2.600000	&	2.603374	&	0.201804	&	1.553823	&	0.358145	&	2.606837	&	0.179346	&	2.382969	&	0.617191	\\
2.0	&	2.700000	&	2.696222	&	0.194625	&	1.628281	&	0.407107	&	2.702147	&	0.182631	&	2.305122	&	0.883883	\\
2.0	&	2.800000	&	2.802244	&	0.198338	&	1.671113	&	0.416296	&	2.818203	&	0.190096	&	2.292552	&	0.999156	\\
2.0	&	2.900000	&	2.904717	&	0.209044	&	1.757733	&	0.452250	&	2.918469	&	0.191426	&	2.191101	&	1.186119	\\
2.0	&	3.000000	&	3.000119	&	0.200248	&	1.794150	&	0.449686	&	3.009131	&	0.205601	&	2.091009	&	1.306820	\\
2.0	&	3.100000	&	3.095304	&	0.199938	&	1.861154	&	0.493821	&	3.120773	&	0.203151	&	2.027416	&	1.389508	\\
2.0	&	3.200000	&	3.206066	&	0.200175	&	1.928786	&	0.495782	&	3.210318	&	0.209492	&	1.922314	&	1.469364	\\
2.0	&	3.300000	&	3.305481	&	0.191043	&	2.017072	&	0.513027	&	3.318200	&	0.220698	&	1.704367	&	1.569541	\\
2.0	&	3.400000	&	3.397078	&	0.198111	&	2.045195	&	0.562483	&	3.406487	&	0.224814	&	1.631980	&	1.610402	\\
2.0	&	3.500000	&	3.510252	&	0.208732	&	2.105675	&	0.588506	&	3.516766	&	0.231254	&	1.439036	&	1.648696	\\
2.0	&	3.600000	&	3.602030	&	0.206381	&	2.150111	&	0.604538	&	3.619306	&	0.233198	&	1.341696	&	1.673103	\\
2.0	&	3.700000	&	3.691687	&	0.201116	&	2.245178	&	0.687834	&	3.713207	&	0.232729	&	1.240939	&	1.677783	\\
2.0	&	3.800000	&	3.814261	&	0.194869	&	2.285464	&	0.694195	&	3.822946	&	0.250042	&	1.173427	&	1.684988	\\
2.0	&	3.900000	&	3.899518	&	0.198059	&	2.332963	&	0.720083	&	3.915391	&	0.253124	&	1.115622	&	1.664230	\\
2.0	&	4.000000	&	4.007378	&	0.208750	&	2.446729	&	0.717167	&	4.025399	&	0.257137	&	0.951611	&	1.650998	\\
2.0	&	4.100000	&	4.108197	&	0.209448	&	2.473150	&	0.788030	&	4.117051	&	0.254371	&	0.863239	&	1.593827	\\
2.0	&	4.200000	&	4.203942	&	0.206663	&	2.561313	&	0.789212	&	4.232547	&	0.274843	&	0.830775	&	1.621492	\\
2.0	&	4.300000	&	4.303620	&	0.199229	&	2.637333	&	0.854930	&	4.325485	&	0.272846	&	0.783813	&	1.597587	\\
2.0	&	4.400000	&	4.399662	&	0.196009	&	2.689233	&	0.854584	&	4.413641	&	0.279349	&	0.608025	&	1.497785	\\
2.0	&	4.500000	&	4.515249	&	0.199041	&	2.720509	&	0.910855	&	4.513320	&	0.282582	&	0.701798	&	1.558721	\\
\hline\hline
\end{tabular}
\end{table}

\end{document}